\documentclass[12pt]{article}
\pdfoutput=1
\usepackage{etex}
\usepackage{epsfig}
\usepackage{epsf}
\usepackage{latexsym}
\usepackage{amsfonts,amssymb,amsmath} 
\usepackage{mathtools}
\usepackage{color}
\usepackage[a4paper,vmargin=3cm,hmargin=2.1cm]{geometry}
\usepackage{multirow}
\epsfverbosetrue
\newcommand{\erf}{\text{\rm{erf\,}}}

\newcommand{\de}{\hbox{\rm{d}}}

\newcommand{\lb}{\left[}
\newcommand{\rb}{\right]}
\newcommand{\lp}{\left(}
\newcommand{\rp}{\right)}

\newcommand{\dpp}{\vcentcolon}
\newcommand{\bb}{\begin{eqnarray}}
\newcommand{\ee}{\end{eqnarray}}
\newcommand{\eee}{\nonumber\end{eqnarray}}
\newcommand{\qq}{\quad}

\begin{document}

\thispagestyle{empty}

\begin{center}
${}$
\vspace{2cm}

{\Large\textbf{On a quadratic equation of state and a universe mildly bouncing above the Planck temperature\footnote{supported by the OCEVU Labex (ANR-11-LABX-0060) funded by the
"Investissements d'Avenir" 
\\\indent\qq
French government program
}}} \\

\vspace{2cm}

{\large
Joanna Berteaud\footnote{Aix Marseille Univ, CNRS/IN2P3, CPPM, Marseille, France
\\\indent\qq
joanna.berteaud@etu.univ-grenoble-alpes.fr},
Johanna Pasquet\footnote{Aix Marseille Univ, CNRS/IN2P3, CPPM, Marseille, France
\\\indent\qq
pasquet@cppm.in2p3.fr },
Thomas Sch\"ucker\footnote{
Aix Marseille Univ, UniversitŽ de Toulon, CNRS, CPT, Marseille, France
\\\indent\qq
thomas.schucker@gmail.com },
Andr\'e Tilquin\footnote{Aix Marseille Univ, CNRS/IN2P3, CPPM, Marseille, France
\\\indent\qq
 tilquin@cppm.in2p3.fr }

}

\vspace{2cm}

\hfill{\em To the memory of Daniel Kastler and Raymond Stora}

\vspace{2cm}

{\large\textbf{Abstract}}

\end{center}

A 1-parameter class of quadratic equations of state is confronted with the Hubble diagram of supernovae
and Baryonic Acoustic Oscillations. The fit is found to be as good as the one using the  $\Lambda $CDM model.  The corresponding universe has no initial singularity, only a mild bounce at a temperature well above the Planck temperature. 

\vspace{3cm}

\vskip 1truecm

\eject

\section{Introduction}

One attempt at easing the persisting tension between cosmic observations and theory is to admit an exotic matter component  with an ad hoc equation of state expressing the pressure $p$ of the component as a function of its energy density $\rho $:
\bb p= f(\rho )\label{baro}.\ee
A popular example is the Chaplygin gas \cite{chap} and its generalisations with $f(\rho )= p_0+ w\,\rho +\alpha /\rho $. Another functional class, quadratic equations of state, $f(\rho )= p_0+ w\,\rho +\alpha \,\rho^2 $, has recently attracted attention, but goes goes back at least to the '90ies:

In 1989 Barrow \cite{barr} solves analytically the Friedman equations for equations of state with $f(\rho )= -\rho +\alpha \,\rho^q $, re-expresses his solutions as coming from scalar fields with particular self-interactions, which induce inflation.

Nojiri \& Odintsov \cite{no} note that for positive $\gamma $ and $\lambda $ the deceleration parameter of the scale factor $a(t)=a_0\,t^\gamma  \exp (\lambda\, t)$ changes sign and asked the question: What equation of state induces this scale factor? Their answer is a particular quadratic equation with $p_0=4(3-1/\gamma )\,\lambda ^2/\kappa ^2 ,\ w=-4/(3\gamma  )$ and $\alpha =-(3-2/\gamma )\kappa ^2/18$.

\v{S}tefan\v{c}i\'c and Alcaniz \cite{sa} take up Barrow's model in the light of future singularities like the big rip.

Using dynamical systems theory
Ananda \& Bruni \cite{ana} classify the many different behaviours of the Robertson-Walker and Bianchi I universes resulting from quadratic equations of state. Linder \& Scherrer \cite{lin} analyse asymptotic past and future evolution of universes with ``barotropic fluids'' i.e. matter with an equation of state (\ref{baro}).  

Motivated by Bose-Einstein condensates as dark matter, Chavanis \cite{chav} has written a very complete series of papers on equations of state with $f(\rho )= w\,\rho +\alpha \,\rho^q $ in Robertson-Walker universes including analytical solutions of the Friedman equations and connections with inflation.

Bamba et al. \cite{odin} classify possible singularities of Robertson-Walker universes in presence of exotic matter, in particular with quadratic equations of state. Adhav, Dawande \& Purandare \cite{adh} and
 Reddy, Adhav \& Purandare \cite{redd} solve the Friedman equations analytically in Bianchi I universes with equations of state $f(\rho )= -\rho +\alpha \,\rho^2 $.
Singh \&  Bishi \cite{Singh} consider the same setting in some $f(R,T)$ modified gravity theories.

 Sharov \cite{sha} confronts Friedman universes with curvature and general quadratic equations of state to the supernovae and baryon acoustic oscillation data. 
 
 We would like to contribute to this discussion and concentrate on the particular case $p_0=0$ and $w=-1$. We justify our choice by Baloo's mini-max principle: a maximum of pleasure with a minimum of effort.
 
We find that an exotic fluid with the quadratic equation of state $p=-\rho +\alpha \,\rho ^2$ is an interesting alternative to dark energy, $p=w\,\rho $. Alone, this exotic fluid fits the Hubble diagram of supernovae as well as $\Lambda $CDM. Its universe has no initial singularity. Instead it features a mild bounce, temperatures below 7.4 K, a maximum redshift of 1.7 and blueshifts with high apparent luminosities. We dub this solution ``cold bounce''.  Adding cold matter to this exotic fluid, and combining supernovae with Baryonic Acoustic Oscillations (BAO), preserves the quality of the fit and the absence of an initial singularity. However it leads to high redshifts and temperatures, even above the Planck temperature, ``hot bounce''.

We also show that adding this same exotic fluid  to an isolated spherical star does not upset the successes of general relativity in our solar system.

\section{Cold bounce}

To start, let us write the Friedman equations without cosmological constant and without spatial curvature; we only include an exotic fluid with density $\rho $ and pressure $p$:
\begin{align}
3\,H^2&=8\pi G\,\rho ,\label{first}\\
2\,H'+3\,H^2&=-8\pi G\,p.\label{second}
\end{align}
The prime denotes derivatives with respect to cosmic time $t$. We set the speed of light to one and have the scale factor $a(t)$ carry dimensions of time. The Hubble parameter  is as usual $H\dpp=a'/a$. We can trade the second Friedman equation (\ref{second}) for the continuity equation,
\begin{align}
\rho '=-3H\,(\rho +p).\label{cont}\end{align} 
The coefficient $\alpha $ in the quadratic equation of state has dimensions and it will be convenient to write it in the form,
\begin{align}
p=-\rho +\alpha \,\rho ^2= -\rho +{\textstyle\frac{1}{4}}\,8\pi G\,\tau^2  \,\rho ^2,\label{statealf}\end{align}
where $\tau$ is a characteristic time.
Using this equation of state, the continuity equation (\ref{cont}) integrates readily:
\begin{align}
\rho =\,\frac{1}{8\pi \,G}\,\frac{3\,H_0^2}{1+{\textstyle\frac{9}{4}}\,H_0^2\tau^2\ln(a/a_0) }\, ,\end{align}
with integration constant $H_0\dpp=H(t_0)$, the Hubble parameter today which is related to the density today and $a_0\dpp=a(t_0)$ is the scale factor today, that without loss of generality can be set to $a_0=1$s in flat universes.
For vanishing $\tau$, we retrieve of course the cosmological constant $\Lambda =8\pi G\,\rho $. 

The pleasure continues and the first Friedman equation (\ref{first}) integrates as easily:
\begin{align}
a=a_0\,\exp\,\frac{(t-t_{\rm cb})^{2/3}-(t_0-t_{\rm cb})^{2/3}}{\tau^{2/3}}\, ,\label{asneeze}
\end{align}
where $t_{\rm cb}$ is another integration constant. This scale factor agrees with the one obtained by Chavanis in appendix A of his second paper in reference \cite{chav}.

\begin{figure}[h]
\begin{center}
\includegraphics[width=14.5cm, height=10.5cm]{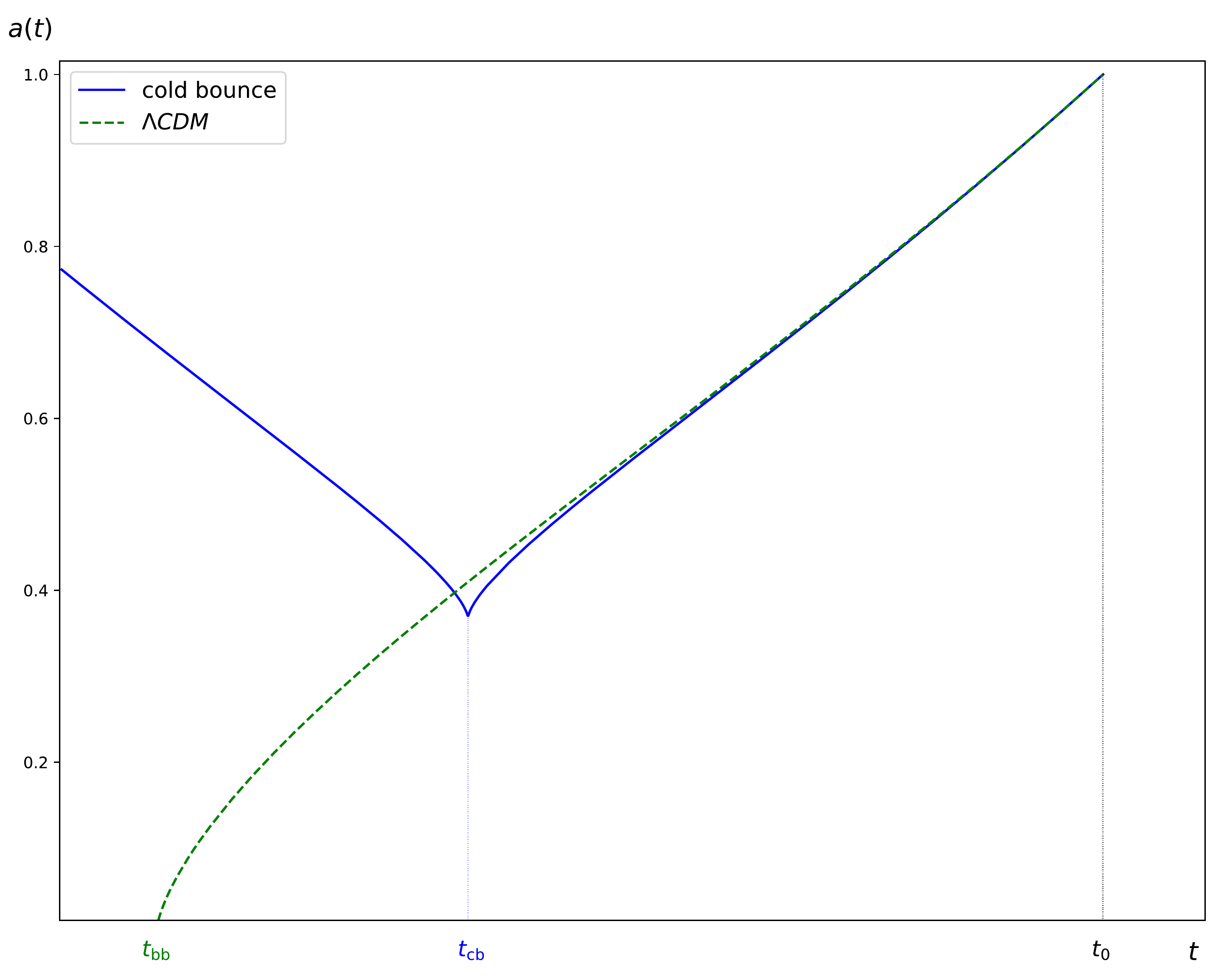}
\caption{Two scale factors, $t_{bb}$ time of big bang,  $t_{cb}$ time of cold bounce, $t_0$ today \label{2a}
 }
\end{center}
\end{figure}

Figure \ref{2a} shows this scale factor together with the one of the $\Lambda $CDM universe,
\begin{align}
a_{\Lambda {\rm CDM}}\,=\,\lp\frac{\cosh[\sqrt{3\Lambda }\,(t-t_{\rm bb})]-1}{\cosh[\sqrt{3\Lambda }\,(t_0-t_{\rm bb})]-1}\rp^{1/3}\, ,
\end{align}
with $t_{\rm bb}$ the time when the big bang occurs.
 The fit is in anticipation of section \ref{versus}.    
 \medskip

 Already now, the scale factor (\ref{asneeze}) tells us a funny story: 
\begin{itemize}
\item no violent, hot bang, just a mild cold bounce at $t=t_{\rm cb}$. By mild we mean that the scale factor remains positive and the temperature finite, only the Hubble parameter diverges together with the density and pressure of our exotic fluid. In the wording of reference \cite{odin} this singularity is of type III.
\item
After the bounce, the universe expands with deceleration until $t_\pm=t_{\rm cb}\,+\,\tau/2^{3/2}$ where the deceleration parameter changes sign. Then the expansion accelerates for ever causing an event horizon.
\item
The bounce is a mild singularity, i.e. integrable and -- if its finite temperature is sufficiently low -- transparent to light so that we can observe its past. There the universe contracts up to arbitrarily negative times, however with a particle horizon.
\item
The scale factor is invariant under time reversal with respect to the time $t_{\rm cb}$.
\end{itemize}
We are sufficiently intrigued by this toy-universe to try and confront it with supernova data.

\section{Hubble diagram and cold bounce}

In the universe of the cold bounce, the Hubble constant is
\begin{align}
H_0=\,\frac{2}{3} \,\frac{1}{\tau^{2/3}(t_0-t_{\rm cb})^{1/3}}\, , 
\end{align}
the redshift is given by
\begin{align}
z+1=\,\frac{a_0}{a}\, =\,\exp-\,\frac{(t-t_{\rm cb})^{2/3}-(t_0-t_{\rm cb})^{2/3}}{\tau^{2/3}},\end{align}
and the apparent luminosity is
\begin{align}
\ell=\,\frac{L}{4\pi \,a_0^2\,\chi ^2}\,\frac{a^2}{a_0^2}\,,  \label{luminosity_bis}  
\end{align}
with the absolute luminosity $L$ and the dimensionless comoving geodesic distance
\begin{align}
\chi(t)\dpp =& \int_t^{t_0}\de\tilde t/a(\tilde t)\\[2mm]
=&\,{\textstyle\frac{3}{2}} \,\frac{\tau}{a_0}\, \exp R^2\lb-x^{1/3}\exp -x^{2/3}+{\textstyle\frac{1}{2}} \sqrt{\pi}\,\erf x^{1/3}\rb_{(t-t_{\rm cb})/\tau} ^{(t_0-t_{\rm cb})/\tau}\\[2mm]
=&\,{\textstyle\frac{3}{2}} \,\frac{\tau}{a_0} \lb
-R +{\textstyle\frac{1}{2}} \sqrt{\pi }\,\exp R^2 \,\erf R+
\exp R^2\,\lp \frac{t-t_{\rm cb}}{\tau}\rp^{1/3}\exp -\lp \frac{t-t_{\rm cb}}{\tau}\rp^{2/3} \right.\nonumber\\
&\qq\qq\qq\left.-{\textstyle\frac{1}{2}} \sqrt{\pi }\,\exp R^2 \,\erf \lp \frac{t-t_{\rm cb}}{\tau}\rp^{1/3}\rb,\label{only}
\end{align}
with the abbreviation
\begin{align} R\dpp=\lp \frac{t_0-t_{\rm cb}}{\tau}\rp^{1/3} = \,\frac{2}{3} \,\frac{1}{\tau H_0}\,  \label{rdef}
\end{align} 
and the error function 
\begin{align}
\erf x\dpp=\,\frac{2}{\sqrt{\pi }}\int_0^x\exp -y^2\,\de y.
\end{align}
Note that the formula (\ref{only}) is a priori valid only for emission times $t\in [t_{\rm cb},t_0]$. However due to the time reversal symmetry of the scale factor with respect to $t_{\rm cb}$, equation (\ref{only}) is valid for all $t$. In particular we have
\bb \chi (t)= 2\chi (t_{\rm cb})-\chi (2t_{\rm cb}-t).\label{reversal}\ee
 \begin{figure}[h]
\begin{center}
\includegraphics[width=14.5cm, height=10.5cm]{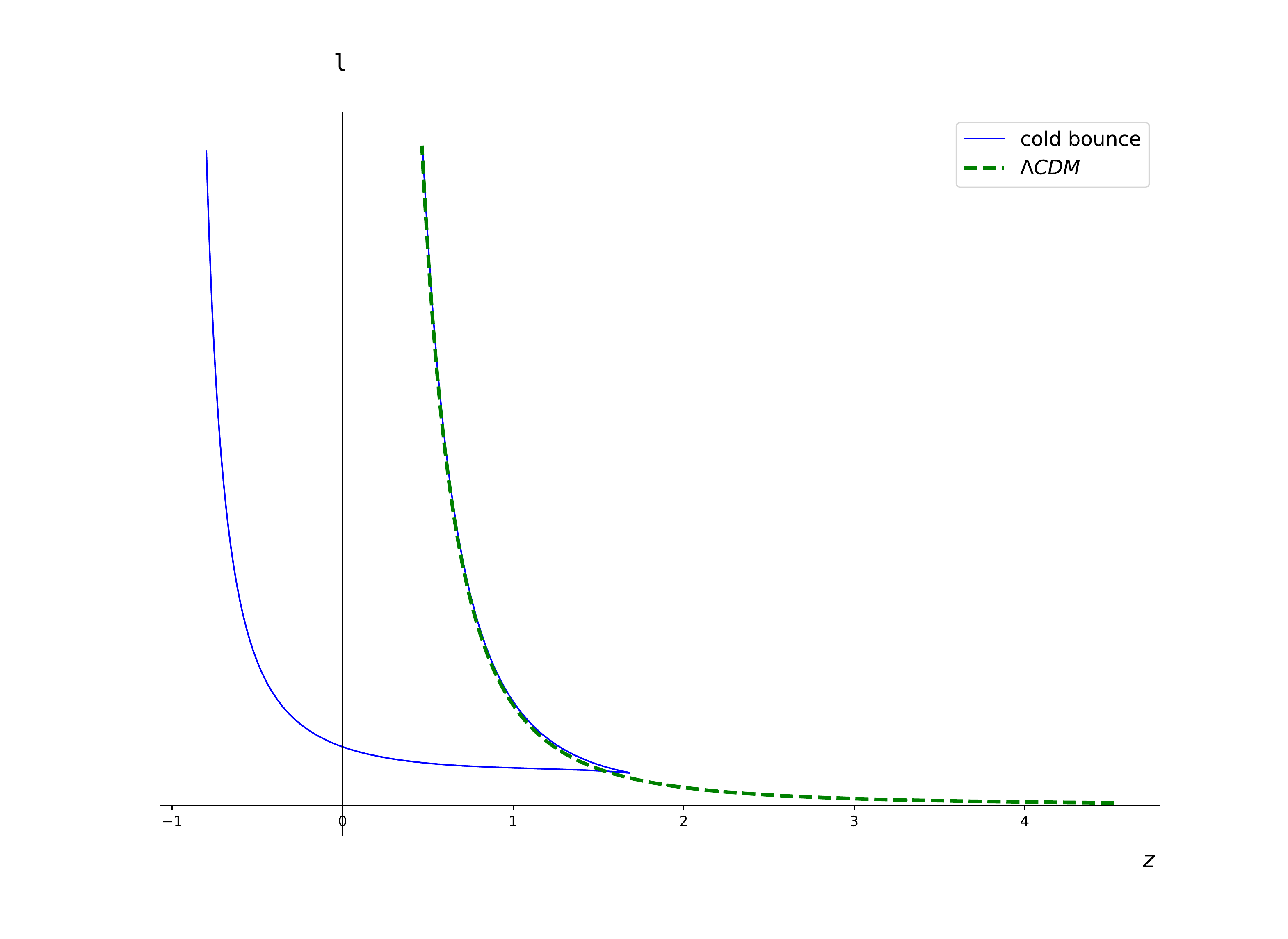}
\caption{Hubble diagrams of cold bounce and $\Lambda $CDM universes
\label{2ell} }
\end{center}
\end{figure}

Note that the redshift is not an invertible function of time, it has a maximum:
\begin{align}
z_{\rm max}=z(t_{\rm cb})=\,\exp\,\frac{(t_0-t_{\rm cb})^{2/3}}{\tau^{2/3}}-1, \label{zmax}
\end{align}
and takes negative values, blueshift, for $t<2t_{\rm cb}-t_0$. Its minimal value $z_{\rm min}=-1$ occurs at $t=-\infty$ corresponding to  the particle horizon of finite comoving geodesic distance
\begin{align}
\chi _{\rm horizon}=\chi (-\infty)
=\,{\textstyle\frac{3}{2}} \,\frac{\tau}{a_0} \lb
-R +{\textstyle\frac{1}{2}} \sqrt{\pi }\,\exp R^2 \,\{\erf R+1\}
 \rb.
\end{align}
At this blueshift, $z_{\rm min}=-1$, the apparent luminosity tends to infinity because the energy of the arriving photon is boosted during the long contraction phase.

Figure \ref{2ell} shows the Hubble diagrams of  cold bounce and  $\Lambda $CDM universes.
Because of the non-invertibility of the function $z(t)$ the Hubble diagram is a genuine parametric plot containing two functions $\ell_{1,2} (z)$,
\begin{align}
\ell_{1,2}(z)=\,\frac{L}{9\pi \, \tau^2}\,\frac{1}{(z+1)^2\,I_{1,2}^2(z)}\, \label{luminosity}.
\end{align}
The function $I_1(z)$ comes from emission times $t\in[t_{\rm cb},\,t_0)$ corresponding to $z\in(0,\,z_{\rm max}]$, 
\begin{align}
I_1(z)=&\int_0^z\sqrt{R^2-\ln (\tilde z+1)}\,\de\tilde z\\
=&\,-R +{\textstyle\frac{1}{2}} \sqrt{\pi }\,\exp R^2\,\erf R\nonumber\\
&+(z+1)\, \sqrt{R^2-\ln ( z+1)}-{\textstyle\frac{1}{2}} \sqrt{\pi }\,\exp R^2\,\erf \sqrt{R^2-\ln ( z+1)}\,\, .\label{iz}
\end{align}
The function $I_2(z)$ comes from emission times $t\in(-\infty,\,t_{\rm cb}]$ corresponding to $z\in(-1,\,z_{\rm max}]$, 
\begin{align}
I_2(z)=\,&2\,I_1(z_{\rm max})-I_1(z)\label{zreversal}
\\
=&\,-R +{\textstyle\frac{1}{2}} \sqrt{\pi }\,\exp R^2\,\erf R\nonumber\\
&-(z+1)\, \sqrt{R^2-\ln ( z+1)}+{\textstyle\frac{1}{2}} \sqrt{\pi }\,\exp R^2\,\erf\sqrt{R^2-\ln ( z+1)}\,\, ,
\end{align}
where equation (\ref{zreversal}) is another manifestation of the time reversal symmetry via equation (\ref{reversal}).

An example of a Hubble diagram with red- and blueshift, but without a cusp associated to the mild bounce can be found in reference \cite{st05}.

\section{Cold bounce versus supernovae}\label{versus}

To analyze the effect of a quadratic equation of state on the evolution of the universe,
we use data sets of type 1a supernovae from the Joint Light curve Analysis \cite{jla} 
with 740 supernovae up to a redshift
of 1.3. The JLA analysis simultaneously fits cosmological parameters including normalization parameter $m_s$
with light-curve time-stretching $\alpha_s$ and colour at maximum brightness $\beta_c$.  
We use frequentist's statistics \cite{pdg} based on $\chi^2$ minimization. 
The MINUIT package  \cite{minuit} is used to find the minimum of the $\chi^2$ and 
to compute errors by using the second $\chi^2$ derivative. All our 
results are given after marginalization over the nuisance parameters ($m_s$, $\alpha_s$ and $\beta_c$).

The general $\chi^2$ is expressed in terms of the full covariance matrix of
supernovae magnitude stretch and colour including correlations and systematics. It reads
\bb \chi^2 = \Delta M^T V^{-1} \Delta M, \ee
where  $\Delta M$  is the vector of differences between the expected supernovae 
magnitude $m_e$ and the reconstructed experimental magnitude at maximum of
the light curve $m_r$. The reconstructed magnitude reads:
\bb m_r = m_{\rm peak} + \alpha_s X1 - \beta_c C, \ee
where $X1$ is related to the measured light curve time stretching, $C$ the supernovae colour 
at maximum of brightness and $m_{peak}$ the magnitude at maximum of the light
curve fit.  

The expected magnitude is written as  $m_e(z) = m_s - 2.5 \log_{10} \ell(z)$ where
$\ell(z)$ is given by the first branch of equation (\ref{luminosity}). 
Notice that the normalization parameters $m_s$ contain the unknown intrinsic
luminosity of type 1a supernovae as well as the $\tau$ parameter. 
The expected magnitude evolution with redshift is then only a function of $m_s$ and $R$ or 
equivalently $\tau H_0 $ (\ref{iz}). 

Table~\ref{table1} presents the results of the fit of the JLA sample 
for a universe with cold bounce and the flat $\Lambda$CDM universe for comparison. 
The quality of the fits are identical for both models even though it is 
marginally better for the cold bounce. This clearly indicates that the 
exotic fluid with the quadratic equation of state (\ref{statealf}) can replace both dark matter and dark energy  in the
supernovae data. However this success involves three problems.

\begin{figure}[ht]
\begin{center}
\includegraphics[width=14cm, height=7cm]{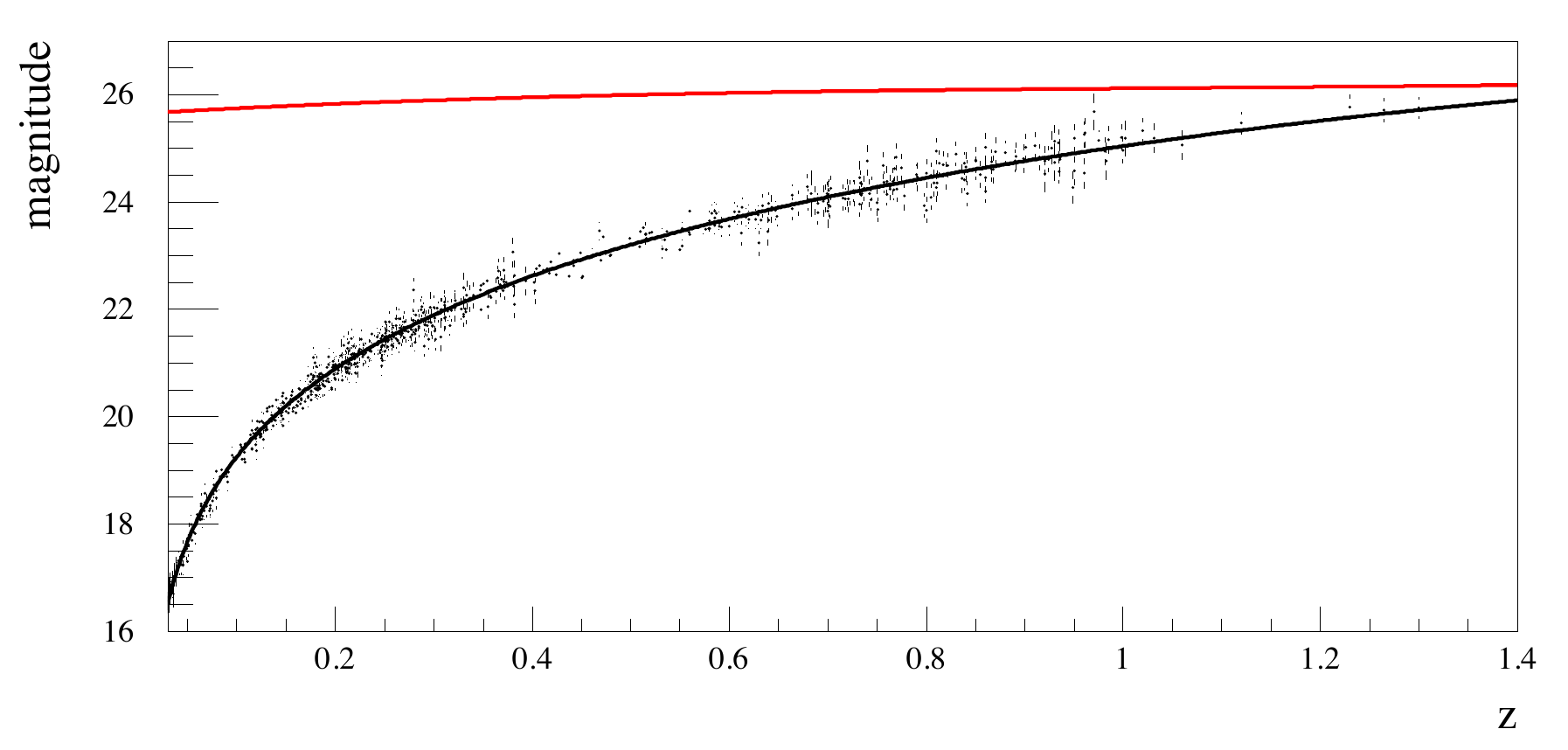}
\caption[]{The Hubble diagram for JLA supernovae. The lower, populated line (in  black) is the first branch of the Hubble diagram computed from the cold bounce (supernovae exploding after the cold bounce), the upper, unpopulated line (in red) is the second branch (supernovae exploding before the cold bounce). Only the black line was used in the fit.}
\label{figure1}
\end{center}
\end{figure}

\begin{table}[htbp]
\begin{center}
\begin{tabular}{||c||c|c|c||c|c||} \hline
            &   \multicolumn{3}{c||}  {cold bounce}  & \multicolumn{2}{c||}{$\Lambda$CDM}  \\ \hline
            &         $\tau H_0 $           &$z_{\rm max}$ &  $\chi^2$            &  $\Omega_{m0} $   &          $\chi^2$    \\ \hline
 JLA        &  $0.66 \pm 0.02$& $1.7\pm 0.3$    &  $739.0$         &  $0.29 \pm 0.03$  & $739.3$  \\ \hline
\end{tabular}
\caption[]{Fit results (1$\sigma$ errors) for JLA
  sample using  the cold bounce and flat $\Lambda$CDM.}
\label{table1}
\end{center}
\end{table}
The first one is that we have discarded the second branch of the Hubble diagram (\ref{luminosity}) since the data are well fitted by only the first branch (see Figure \ref{figure1}). The second problem is the maximum redshift value that we obtained. Indeed as the parameter $\tau H_0$  is equal to about $2/3$ with an error of $1 \,\%$, i.e. $R\sim 1$,
and using formulae (\ref{rdef}) and (\ref{zmax}), the maximum redshift is about 1.7, $z_{\rm max}=e-1$. Consequently the maximum temperature of this universe is  
\bb T_{\rm cb}=(z_{\rm max}+1)\, T_{\rm CMB}=
2.72\cdot2.72\,{\rm K}=7.4\,{\rm K},\ee
at the mild bounce, which took place some 9.1 Giga years ago and which does deserve the name ``cold bounce''. 
Finally the the cold bounce predicts blueshifted supernovae with high apparent luminosities.

\section{Adding  cold matter to the cold bounce \label{adding}}

In this section we show that adding cold matter to the cold bounce makes it hot and solves the above three problems without spoiling its success with supernovae data. We may even add photons and then include BAO data.

The punch line of our approach is as follows: 
\begin{itemize}\item
If we fill a flat universe with the exotic fluid of quadratic equation of state only and we fit this universe to the supernovae data, we obtain the cold (mild) bounce with a quality as good as $\Lambda $CDM.
\item
If we add in this universe cold matter and a pinch of photons to the exotic fluid and add BAO data to the supernovae data, we obtain two bounces, a warm bounce and a hot bounce, both are mild and of good quality.
\end{itemize}
We denote by $\rho _m$ the mass density of cold matter.  Its pressure is zero. Likewise we write $\rho _\gamma $ for the energy density of the photons and $p_\gamma =\rho _\gamma  /3$ for their pressure.
As we now have only one continuity equation,
\begin{align}
(\rho+\rho _m+\rho _\gamma ) '=-3H\,(\rho +\rho_m +\rho _\gamma  + p +p_\gamma ),\label{cont2}\end{align} 
 for three components, our system of differential equations is under-determined. To remain in business we cheat 
as is tradition by postulating three independent continuity equations for the three components. These three continuity equations then integrate like charms yielding:
 \begin{align}
 \Omega _X&\dpp=\,\frac{8\pi G\,\rho }{3\,H_0^2}\,=\,\frac{1}{{\textstyle\frac{9}{4}}\,H_0^2\,\tau^2\,\ln(a/a_0)+1/\Omega _{X0}}\, , \label{omegax_evol}
\\[2mm]
\Omega _m&\dpp=\,\frac{8\pi G\,\rho_m }{3\,H_0^2}\,= \Omega _{m0}\lp\frac{a_0}{a}\rp^3\, ,  \label{omegab_evol}
\\[2mm]
\Omega _\gamma &\dpp=\,\frac{8\pi G\,\rho_\gamma  }{3\,H_0^2}\,=\Omega _{\gamma 0} \lp\frac{a_0}{a}\rp^4. \label{omega_gama_evol}
\end{align} 
From the first Friedman equation (\ref{first}) we have for the initial conditions
  \begin{align}
  \Omega _{X0}+\Omega _{m0}+\Omega _{\gamma 0 } =1. \label{initial}
  \end{align}
  Now remains the integration of the first Friedman equation (\ref{first}) with three components. The exotic fluid alone generates the mild singularity of the cold bounce, while cold matter and photons alone generate the violent singularity of the popular hot big bang. 
  
   \begin{figure}[h]
\begin{center}
\includegraphics[width=14.5cm, height=10.5cm]{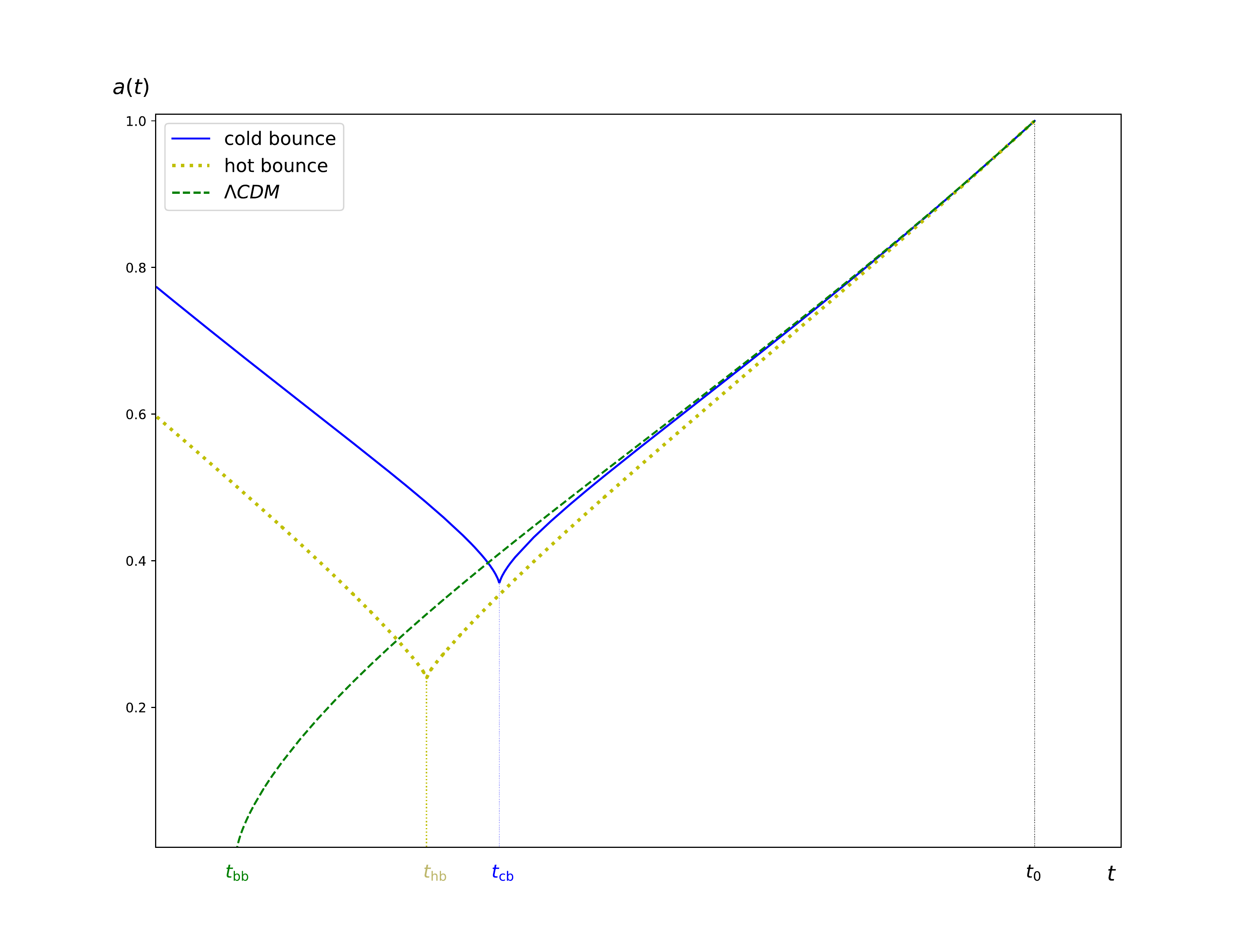}
\caption{Three scale factors,\label{3a} $t_{bb}$ time of big bang,  $t_{hb}$ time of hot bounce, $t_{cb}$ time of cold bounce, $t_0$ today. Attention: in Figure 1  we displayed the scale factor  of the cold bounce with the value $\tau H_0$ obtained later in Section 4 from the observational fit. Here, for graphical reasons, the ``hot bounce'' does not reflect the fitted values from section 6, neither hot nor warm.
 }
\end{center}
\end{figure}
\begin{figure}[h]
\begin{center}
\includegraphics[width=12.5cm, height=10.5cm]{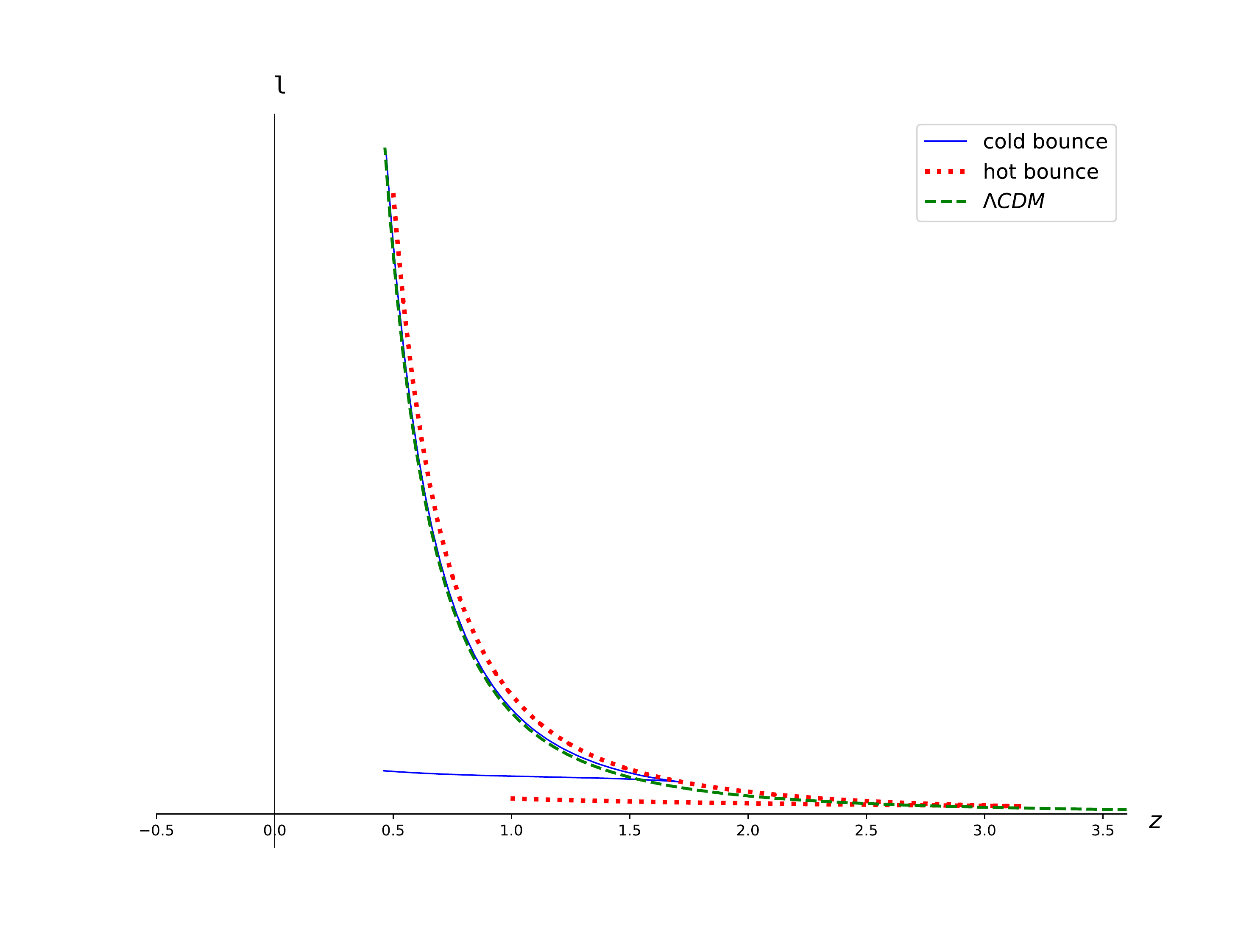}
\caption{Three Hubble diagrams, redshifts only.
 Attention: in Figure 2  we displayed the Hubble diagram of the cold bounce with the value $\tau H_0$ obtained later in Section 4 from the observational fit. Here, for graphical reasons, the ``hot bounce'' does not reflect the fitted values from section 6, neither hot nor warm. \label{3ell}
 }
\end{center}
\end{figure}
  
  With these three components, the first Friedman equation has still separate variables and tells us that the mild cold bounce kills the violent hot big bang:
\begin{align}
  \,\frac{\de a}{a\,\sqrt{[(9/4)\,H_0^2\,\tau^2\,\ln(a/a_0)+1/\Omega _{X0}]^{-1}\,+\,\Omega _{m0} ({a_0}/{a})^3
\,+\,\Omega _{\gamma  0} ({a_0}/{a})^4}}\, =\,H_0\,\de t.\label{sep}
\end{align}
  Indeed going back into the past starting from $t_0$, the scale factor decreases until -- at $t_{\rm hb}$ -- the bracket
$  [(9/4)\,H_0^2\,\tau^2\,\ln(a/a_0)+1/\Omega_{X0}]$ vanishes. At this time, the scale factor has the value
 \begin{align}
a_{\rm hb}\dpp=a(t_{\rm hb}) =a_0\,\exp-\frac{4}{9\,\Omega_{X0}\,H_0^2\tau^2}\, ,
\end{align}
 and the universe bounces mildly. This is how it avoids the big bang. The maximum redshift,
 \begin{align}
z_{\rm max}=\,\frac{a_0}{a_{\rm hb}}\,-1 =\exp\frac{4}{9\,\Omega_{X0}\,H_0^2\tau^2}\,-\,1,
\end{align}
  increases with respect to its value of the cold bounce.
We performed the integration on the left-hand side of equation (\ref{sep}) numerically.
   
   Figure \ref{3a} shows the scale factors for  $\Lambda $CDM, the cold bounce and a hot bounce and Figure \ref{3ell} shows the three corresponding Hubble diagrams, redshifts only.  
   
   Note that by the same mechanism, the bounce continues to kill the big bang when we add spatial curvature.

      \section{Hot bounce versus supernovae and BAO}\label{addbao}

We use the same kind of analysis as in Section \ref{versus} to study the mildly bouncing 
universe now with cold matter and photons added. As we have seen in the previous section,
the evolution of exotic, cold matter and photon densities in terms of the scale factor are
obtained from continuity equations and  given by formulae (\ref{omegax_evol}),(\ref{omegab_evol}) and (\ref{omega_gama_evol}). 

The first Friedman equation (\ref{first}) is solved with these three independent 
components using the Rugge-Kutta algorithm \cite{rk} with a step in time 
corresponding to an equivalent step in redshift well below the 
experimental redshift error ($10^{-3}$) for supernovae analysis and small enough to explore accurately
the scale factor evolution up to the bounce at high redshift.

To speed up the processing and to avoid numerical instability we use a Multi Layer Perceptrons (MLP) neural network \cite{frank,neural} 
with 2 hidden layers of 25 neurons each trained on a grid of cosmological parameters today $\Omega_{m0} , \tau H_0$ and at a fixed value of
$\Omega_{\gamma 0} = 5.38 \cdot 10^{-5}$ \cite{planck}.

The  apparent luminosity $\ell(z)$ of a supernova is computed using equation (\ref{luminosity_bis})  
and by assuming again that we only see supernovae that exploded after the bounce.  We use the closure relation (\ref{initial}) for a flat universe. 

The final fit procedure is then only a function of $m_s$, 
time-stretching correction $\alpha_s$, colour correction $\beta_c$,
$\Omega_{m0}$ and the new dimensionless parameter $\tau H_0$ describing 
the weight of the quadratic term in the equation of state (\ref{statealf}). 


Table \ref{table2} shows the results of the fit for supernovae marginalized over $m_s$, $\alpha_s$ and $\beta_c$.

 Despite the fact that we have one more free parameter, 
the quality of the fit is only marginally improved compared to the cold bounce.  
As before the maximum redshift is still quite small with a value of $z_{\rm max}= 7.5$, but with an infinite error.
 
To compare the $\Lambda$CDM fit in Table 1 with the fit to the bouncing model with cold matter in Table 2, we use the log likelihood ratio theorem \cite{w} by Wilks from 1938, which shows that the $\chi ^2$ differences between both models follow a $\chi ^2$ distribution with a number of degrees of freedom equal to the difference between the degrees of freedom of both models, one in this case. The $\chi ^2$ difference between both models is equal to 0.35 and leads to the conclusion that the bouncing  model with cold matter is as good or better than $\Lambda$CDM at a probability level of 43 \%.

\begin{figure}[ht]
\begin{center}
\includegraphics[width=16.cm, height=13.cm]{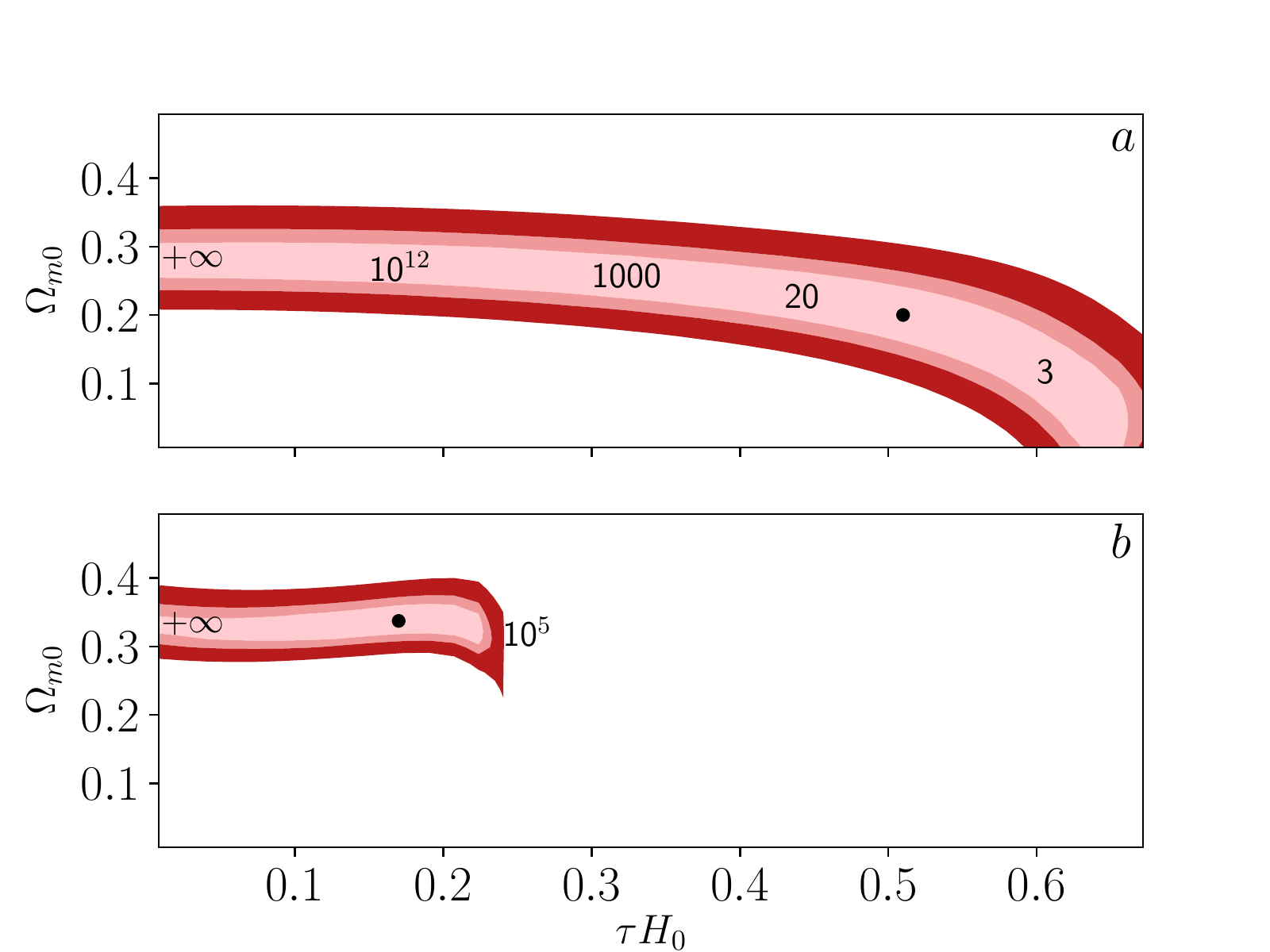}
\caption[]{$68\%$, $95\%$ and $99\%$ confidence levels contours for the bouncing universe with cold matter added. 
  Figure (a) is for supernovae and (b) for BAO. The black dots represent the maxima of probability, and the black
numbers give the maximum redshift $z_{max}$.}
 
\label{figure2}
\end{center}
\end{figure}

Figure \ref{figure2}-a shows the confidence level contour on $\Omega_{m0}$ versus $\tau H_0$. As expected, the degeneracy between 
both fluids is important and explains the big errors in Table \ref{table2}.

To break this degeneracy we choose to use Baryonic Acoustic Oscillations from the last SDSS III data release \cite{sdss3}. Using
361762 galaxies at effective redshift of $z_{\rm LOWZ}=0.32$  and 777202 galaxies at effective redshift of $z_{\rm CMASS}=0.57$, 
the SDDS III collaboration extracts the BAO-scale from the angular direction and radial projection and provides
the dimensionless reduced parameter $R_{\rm BAO} = D_V(z)/r_s(z_d)$ where $D_V$ is the 3-dimensional distance: 
\begin{align}
  D_V(z) = \lp d^2_{\chi}(z) \frac{cz}{H(z)} \rp^{1/3} \, {\rm with} \,\,\, 
  d_{\chi}(z) =  \int_0^z \frac{c}{H(\tilde z)}  \de\tilde z
\end{align}

and $r_s$ the sound horizon at drag epoch $z_ d$:

\begin{align}
r_s(z_d) =  \int_{z_d}^{z_{max}} \frac{v_s (\tilde z)}{H(\tilde z)}  \de\tilde z. \label{rs}
\end{align}

The sound speed reads \cite{sound}:
\begin{align}
v_s(z) = \frac{c}{\sqrt{3\lp 1+\frac{3\, \Omega_b(z)}{4\, \Omega_{\gamma}(z)}\rp}}
\end{align}
 where $\Omega_b$ is the reduced baryonic matter energy density equal to 0.04 today and $\Omega_{\gamma}$ the reduced photon energy
density equal to $5.38 \cdot 10^{-5}$ today \cite{planck}.

The redshift at the drag epoch $z_ d$ is well approximated by \cite{zdrag}:
\begin{align}
z_d &= \frac{1291 \lp \Omega_{m0}\,h^2 \rp^{0.251}}{1+0.659\lp \Omega_{m0}\, h^2 \rp^{0.828}}\lp 1+b_1(\Omega_{b0}\, h^2)^{b_2}\rp \, , \label{z_drag}
\\[2mm]
b_1 &=  0.313 (\Omega_{m0}\, h^2 )^{-0.419} \lp 1+0.607(\Omega_{m0}\, h^2)^{0.674} \rp \, ,  
\\[2mm]
b_2 &= 0.238 \lp \Omega_{m0}\, h^2 \rp^{0.223} \,,
\end{align}
with $h=H_0/100 = 0.677$ \cite{planck}.

We compute the dimensionless parameter $D_V(z)/r_s(z_d)$ at both effective redshift numerically using
our Neural Network function of cosmological parameters $\Omega_{m0} \, , \tau H_0$ and $z$.
The $\chi^2_{\rm BAO}$ reads: 
\begin{align}
 \chi^2_{\rm BAO} = \sum_{i={\rm LOWZ,\,CMASS} } \lp \frac{R_{\rm th}(z_i,\Omega_m^{(0)},\tau H_0)-R_{\rm BAO}(z_i)}{\sigma_i}\rp^2
\end{align}
with $R_{\rm BAO}(z_{\rm LOWZ}) = 8.61 \pm 0.15 $ and  $R_{\rm BAO}(z_{\rm CMASS}) = 13.71 \pm 0.13 $ \cite{sdss3}.

The result of the BAO fit is shown in line 2 of table \ref{table2} and the probability contour in Figure \ref{figure2}-b.
The constraint on $\tau H_0 = 0.17^{+0.06}_{-0.17} $ is tighter than using JLA but still compatible with zero at 1 sigma level.
The maximum redshift of the order of $1.3 \cdot 10^{10}$ is well above the redshift at drag epoch computed with (\ref{z_drag}), 
$z_d = 1020.9$, which is the condition to compute the sound horizon $ r_s $ with equation (\ref{rs}). Figure \ref{figure2}-b clearly
indicates that the main constraint on the bouncing universe is from the size of the sound horizon.

\begin{table}[htbp]
\begin{center}
\begin{tabular}{||c||c|c|c|c|c|c||} \hline

            & $\Omega_{m0}$ &  $\tau H_0$ &$T_{\rm bounce}/{\rm K} $  &   $\chi^2$    \\ \hline
 JLA        & $0.2^{+0.15}_{-0.14}$&$0.51^{+0.08}_{-0.5}$ & $ 9.6 < 23.0 < \infty $  &  $738.95$   \\ \hline
 BAO        & $ 0.34 \pm 0.03  $& $0.17^{+0.06}_{-0.17}$ &$ 9.2\cdot 10^{5}<3.6 \cdot10^{10}< \infty$ & $0.21$  \\ \hline
\multirow{2}{*}{SN+BAO}      &$ 0.28 \pm 0.03 $ & $0.231^{+0.001}_{-0.017}$ & $ 2.6\cdot 10 ^{5}<2.9 \cdot 10^{5} <1.9\cdot 10^{6} $ & $739.11$ \\
                             &$ 0.31 \pm 0.01 $ & $0.08 \pm 0.06$ & $ 5.0\cdot10^{14}<1.4 \cdot10^{44} < 10^{1610} $ & $740.47$ \\ \hline
\multirow{2}{*} {LSST+EUCLID} & $0.28 \pm 0.005$ & $0.231 \pm 0.001$ & $2.6\cdot 10^{5}<2.9\cdot 10^{5}<3.2\cdot 10^{5} $ & N/A \\
                              & $0.31 \pm 0.003$ & $0.08^{+0.03}_{-0.04}$ & $ 3.5\cdot 10^{23}<1.4\cdot 10^{44} <10^{175}$ & N/A \\ \hline
\end{tabular}
\caption[]{Fit results (1$\sigma$ errors) for JLA, BAO, combined and EUCLID+LSST simulations using the bouncing universe with  cold matter added to the exotic fluid. For better readability the bounce temperatures and their lower and upper 1$\sigma$ limits are indicated by inequalities. }
\label{table2}
\end{center}
\end{table}

Since both probes are statistically compatible (Figure \ref{figure2}) we combine them using:
\begin{align}
  \chi^2_{\rm total} (\Omega_{m0},\tau H_0,m_s,\alpha_s,\beta_c)= \chi^2_{\rm SN}(\Omega_{m0},\tau H_0,m_s,\alpha_s,\beta_c)+\chi^2_{\rm BAO}(\Omega_{m0},\tau H_0).
\end{align}

\begin{figure}[ht]
\begin{center}
\includegraphics[width=14.5cm, height=9cm]{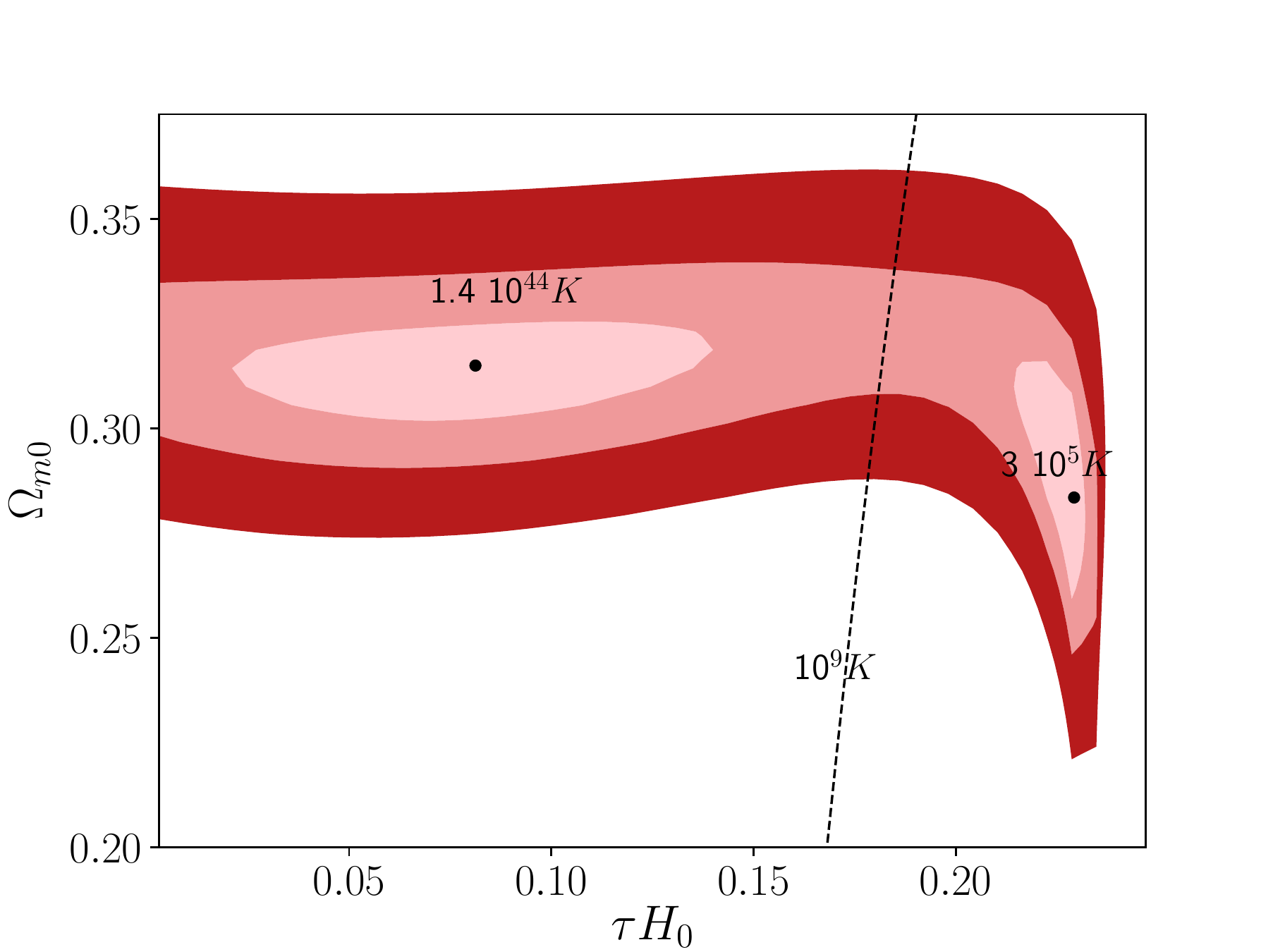}
\caption[]{$68\%$, $95\%$ and $99\%$ confidence levels contours for the mildly bouncing universe with cold matter
 using supernovae from JLA and BAO from SDSS-III. The black numbers represent the temperature of the bounce.
The two black dots represent the two maxima of probability, 
the hot bounce on the left-hand side and the warm bounce on the right-hand side.
Parameters to the left of the dashed black line are compatible with nucleosynthesis.
The full black line indicates the Planck temperature.}
 
\label{figure3}
\end{center}
\end{figure}

The total $\chi^2$ is minimized over all parameters and marginalized over $m_s,\alpha_s$ and $\beta_c$ to obtain the results quoted
in line 3 of Table \ref{table2} and to construct the probability contour shown in Figure \ref{figure3}. We observe two distinct
minima statistically very close: $\Delta \chi^2 = 1.4$.  The primary minimum is at $\tau H_0 = 0.23^{+0.001}_{-0.017}$ corresponding
to a bounce at a redshift of $1.1 \cdot 10^5$ and a temperature of $3 \cdot 10^{5}$ K well below the expected  nucleosynthesis 
temperature at $T = 10^{9}$ K  (dotted line in Figure \ref{figure3}). We dub this bounce ``warm''. The secondary minimum at $\tau H_0 = 0.08 \pm 0.06$ 
is at a redshift of $5 \cdot 10^{43}$ and a temperature of $1.4 \cdot 10^{44}$ K $\sim1.4 \cdot 10^{31}$ GeV well above the Planck temperature 
\bb
T_{\rm Planck}\dpp=
\sqrt{\frac{\hbar \,c^5}{G\,k_{\rm B}^2}} =
 1.4 \cdot 10^{32} \  {\rm K}\ \sim\ 1.2\cdot 10^{19}\ {\rm GeV}.
\ee
We dub this bounce ``hot''.

 In reference \cite{sha} Sharov  uses a completely different parametrisation orthogonal to our parametrisation. He
chooses a negative value of the quadratic dependency ($\beta$ in his notation) by adding a penalty contribution
to his $\chi^2$ to prevent the bounce singularity. On the contrary and by construction in 
equation (\ref{statealf}) we choose a positive value for the quadratic dependency: $\alpha =\beta=\frac{3}{4} (\tau H_0)^2$.
The ensuing mild bounce singularity is supported by supernova and BAO data.

To evaluate the future improvement with LSST \cite{lsst} and EUCLID \cite{euclid} large survey we use a fast and simple simulation.
We simulate 10000 supernovae up to a redshift of 1 with an intrinsic magnitude error  of 0.12 and a photometric redshift error
of $\sigma_z = 0.01(1+z)$ propagated to the magnitude error. For EUCLID we simulate 10 redshift bins from 1 to 2 with the 
corresponding EUCLID statistic of 5 millions galaxies per bin and the error of the reduced parameter scaled down by
 increased statistic.

 We choose alternatively both minima as fiducial cosmology to compute the expected errors 
quoted in the last 2 lines of Table \ref{table2}. We find an improvement of the error on $\tau H_0$ for the warm bounce of at least a factor 10. The main reason is that degeneracy between supernovae and BAO are orthogonal. 
Therefore the combination of LSST and EUCLID will allow us to confirm or rule out the warm bounce without ambiguity.

On the other hand the error on the hot bounce is only slightly improved because the degeneracies of these probes are
aligned with the $\tau H_0$ axis.

Finally we estimate the expected sensitivity to constrain the bounce. To this end we use the $\Lambda$CDM fiducial cosmology 
to simulate LSST and EUCLID and find $\tau H_0 < 0.14$ at
$95 \% $ confidence level corresponding to a bounce temperature of $T_{\rm bounce}>3.5 \cdot10^5$ K.

\section{Exotic fluid and Schwarzschild solution}\label{schw}

One of the remarkable properties of general relativity is that it describes gravity from scales of the falling apple all the way up to cosmological ones. The short scales invoke the Schwarzschild solution, the long scales rely on the cosmological principle and the Friedman solution. Today's tests of general relativity in the solar system have a precision at the $10^{-5}$ level. On cosmological scales we are at the 10 \% level and various tensions between observation and theory motivate the introduction of an exotic fluid, for example with  equation of state 
\bb 
p=-\rho +\gamma \,\lp\frac{\rho }{\rho _0}\rp^{q-1}\,\rho.\label{stategam} \ee 
 Certainly the most popular exotic fluid has negative pressure, $\gamma =0$. The popularity of this fluid has a simple explanation: its addition is equivalent to the addition of a cosmological constant in Friedman's equations (if we accept the cosmological principle). In this scheme however, the cosmological constant is degraded from a universal constant (akin to Newton's constant) in Einstein's equations to an initial condition (akin to today's exotic fluid density) in Friedman's equations, $\Lambda =8\pi \,G\,\rho _0$.

For nonvanishing $\gamma $, when the exponent is negative, the fluid is a generalised
 Chaplygin gas \cite{chap}. For $q=1$, it is dark energy and it is often used to fit observation: $\gamma  =\dpp w+1\sim0\pm0.1$ after adding a fair amount of dark matter. As we have seen above, the quadratic equation of state, $q=2$,  after adding a fair amount of dark matter, produces a good fit to the Hubble diagram and to the BAO with 

 \bb 
 \gamma =\dpp\,\frac{1}{4}\, \Lambda \tau^2=\,\frac{3}{4}\, \Omega_{X0} (\tau H_0)^2\sim 0.003, 
\ee
for the hot bounce.
Of course we are afraid that the presence of this exotic fluid upsets the mentioned remarkable property of general relativity. We therefore compute the modifications that the exotic fluid induces in the exterior Schwarzschild solution.

\subsection{Static, spherical solutions}

The cosmological principle postulates a six dimensional isometry group. In the case of a static, spherical star the isometry group is only four dimensional.
Therefore the most general metric tensor contains two positive functions $B(r)$ and $A(r)$ (whereas the most general metric allowed by the cosmological principle only contains one function, the scale factor $a(t)$),
\bb \de\tau^2=B\,\de t^2-A\,\de r^2-r^2\,\de\theta ^2-r^2\sin^2\theta \,\de\varphi ^2 ,\ee
and the most general energy-momentum tensor contains three functions of the radius, the energy density $\rho $, the radial pressure and the angular pressure. In order to be able to make sense of the equation of state, we simply assume that both pressures are equal and denote them by $p$.

Let us note that a `polytropic' equation of state 
\bb
p= \gamma \,\lp\frac{\rho }{\rho _0}\rp^{q-1}\,\rho,
\ee
has been used by Tooper \cite{tooper} (in his thesis directed by Chandrasekhar) in order to model the interior metric of a static, spherical star made of `polytropic matter'. Today the polytropic  equation is still used to model relativistic stars. In the following we will be concerned with the metric outside the star. Because of the high symmetry, we do not need to know what matter constitutes the star. However we will suppose that the star is embedded in the exotic fluid with equation of state (\ref{stategam}).

The $tt$, $rr$ and $\theta \theta $ components of Einstein's equation with vanishing cosmological constant can be written:
\begin{align}
\,\frac{A'}{rA^2}\, +\,\frac{1}{r^2}\, \lp 1-A^{-1}\rp&=8\pi \,G\,\rho ,& \\[2mm]
\,\frac{B'}{rAB}\, -\,\frac{1}{r^2}\, \lp 1-A^{-1}\rp&=8\pi \,G\,p, &\\[2mm]
\,\frac{1}{2}\,\frac{B''}{AB}\,  -\,\frac{1}{4}\,\frac{1}{AB}\,\lp\,\frac{A'}{A}\, +\,\frac{B'}{B}\, \rp-\,\frac{1}{2}\,\frac{1}{rA}\,\lp\,\frac{A'}{A}\, -\,\frac{B'}{B}\, \rp &=8\pi \,G\,p, &
\end{align}
where in this section the prime stands for a derivative with respect to $r$. It is convenient to use the equivalent system: the $tt$ component, the sum of the $tt$ and $rr$ components and the covariant energy-momentum conservation:
\begin{align}
\lp r-\,\frac{r}{A}\,\rp'&= 8\pi \,G\,r^2\rho ,\label{tt} &\\[2mm]
\,\frac{1}{rA}\,\lp\,\frac{A'}{A}\, +\,\frac{B'}{B}\, \rp&=8\pi \,G\,(\rho +p), &\\[2mm]
p'+\,\frac{1}{2}\,\frac{B'}{B}\, (\rho +p)&=0.&\label{DT}
\end{align}
Eliminating the pressure with the equation of state (\ref{stategam}) we have three first order equations in three unknowns $A,\,B$ and $\rho $. 

The popular fluid with $p=-\rho $ is again equivalent to allowing for the cosmological constant $\Lambda =8\pi \,G\,\rho _0$. The complete solution is the exterior Kottler (or Schwarzschild-de Sitter) solution:
\bb B=1-{S}/{r}-{\textstyle\frac{1}{3}} \, \Lambda \,r^2=\dpp B_0,\qq
A=B_0^{-1}=\dpp A_0,\qq \rho =\rho _0,\qq p=-\rho _0,\ee 
with integration constants $S$ (the Schwarzschild radius) and $\rho _0$.

For the quadratic equation of state with positive $\gamma  $, one can solve the system (\ref{tt}\,-\,\ref{DT}), but the solution contains the roots of a third order polynomial and is quite complicated. We prefer to solve the system for generic exponent $q$ to first order in $\gamma  $. Let us set
\bb B(r)\sim B_0(r)\,[1+\gamma  \,b(r)],\qq
A(r)\sim A_0(r)\,[1+\gamma  \,a(r)],\qq
\rho (r)\sim \rho _0\,[1+\gamma  \,c(r)].\ee
Plugging this ansatz into the system (\ref{tt}\,-\,\ref{DT}) we find to first order in $\gamma  $:
\bb 
\lp \,\frac{r}{A_0}\, a\rp '\sim\Lambda \,r^2\,c,\qq
\,\frac{1}{rA_0}\,(b'+a')\sim \Lambda ,\qq
 c\sim {\textstyle\frac{1}{2}} \ln B_0\, +\, K_1,\ee
 with an integration constant $K_1$.
 
 Note that the exponent $q$ does not contribute in first order and our analysis also applies to dark energy with $w$ constant and sufficiently close to $-1$.
 
Let us restrict the radius $r$ to values such that $S/r$ and ${\textstyle\frac{1}{3}} \Lambda r^2$ are small enough to justify a linearization also in these two variables. We now denote by $\sim$ the linearization in $\gamma  ,\,S/r$ and ${\textstyle\frac{1}{3}} \Lambda r^2$.
With $\ln B_0\sim-S/r- {\textstyle\frac{1}{3}} \Lambda r^2$ we obtain:
\bb a\sim \,\frac{ K_2}{r}\, ,\qq
b\sim K_3 -\,\frac{ K_2}{r}\, ,\qq
c\sim K_1,\ee
with two more integration constants $K_2$ and $K_3$. Therefore we have to first order:
\begin{align}
B&\sim(1+\gamma K_3)\lb 1-\,\frac{S+\gamma K_2}{r}\, -\,\frac{1}{3}\,\Lambda r^2\rb,\\[2mm]
A&\sim1 +\,\frac{S+\gamma K_2}{r}\, +\,\frac{1}{3}\,\Lambda r^2,\\[2mm]
\Lambda &\sim8\pi \,G\,(\rho _0+\gamma K_1).
\end{align}

\subsection{Robustness of Kottler's solution}

By the coordinate transformation $\tilde t=(1+\gamma K_3/2)\,t$ we can dispose of $K_3$, by a renormalization of the product -- Newton's constant times mass of the central star -- we can dispose of $K_2$ and by a renormalization of the cosmological constant we can dispose of $K_1$. Therefore to first order, the addition of exotic fluids with equations of state (\ref{stategam}) do not modify the outer Kottler solution. 

We are surprised by the robustness of the Kottler solution with respect to small modifications coming from exotic equations of state (\ref{stategam}), because the $\Lambda $CDM solution does change dramatically under the same small modifications with $q=2$: from the big bang to a mildly bouncing universe.
In other words: the limit as $\tau$ goes to zero of the age of the universe is not continuous.

The robustness of the Kottler solution is comparable to the robustness of this mildly bouncing solution, $\tau \not=0$, with respect to the addition of cold matter, photons and curvature as remarked at the beginning of Section \ref{adding}. 

For those who believe in exotic fluids, the robustness of the Kottler solution may be a relief, because the wished for modifications on cosmological scales do not upset the successful tests of general relativity in our solar system. For us it is a disappointment, we would have  preferred that the proposed modifications be constrained by existing experimental facts. 

In any case, before asking how exotic fluids modify Newton's law of gravitational attraction and the bending of light, we should ask seriously how a cosmological constant modifies these two fundamental phenomena on cosmological scales. Indeed we would be happy to see large-structure simulations, that include the long-range repulsive force coming from a positive cosmological constant. We would also be happy to see the controversy resolved whether the cosmological constant affects strong lensing \cite{contro}.

\section{Conclusions}

Dark energy with constant $w$ is a 1-parameter extension of the successful $\Lambda $CDM model. We have studied another 1-parameter extension with parameter $H_0 \tau$, the ``hot mild bounce''. Let us summarize its salient properties:
\begin{itemize}\item
As in the dark energy model, the cosmological constant is not introduced in the Einstein equation, but is induced by the exotic fluid in a spacetime of high symmetry (cosmological principle or hypothesis of an isolated, static, spherical star).
\item
The hot bounce fits today's supernovae and BAO data as well as $\Lambda $CDM and could be falsified by LSST and EUCLID in a near future.
\item
It replaces the big bang by a mild bounce.
\item
Its bounce occurs at a redshift above $110\,000$ and at a temperature well above the Planck temperature. Therefore the second branch of the Hubble diagram, see Figure \ref{2ell}, with its potentially problematic redshifts and blueshifts are screened by the primordial plasma and arguably also by quantum fluctuations. 
\item
Its nucleosynthesis is unmodified with respect to the standard model.
\item
The exotic fluid modifies the tests of general relativity in our solar system. However these modifictions  are small and do not upset the tests.
\end{itemize}

It is interesting to note that loop quantum cosmology favors a bouncing universe without the need of an exotic fluid \cite{bojo}.

Figure \ref{figure3}  tells us that supernovae and baryonic acoustic oscillations do not segregate between 
hot and warm bounces and the latter does upset nucleosynthesis.  
In order to separate the two bounces, other probes are necessary.
Thanks to the robustness of the Kottler solution presented in Section \ref{schw}, weak lensing data can be used to test the hot bounce model.  Also a confrontation of the hot bounce with Cosmic Micro-wave Background data would certainly be interesting. Its redshift epoch, $ z_{\rm CMB}=1090$, is comparable to the drag epoch of BAO,  $z_d = 1020.9$, cf Section 6, but the CMB analysis is more involved. \\

${}$\vspace{6mm}\\
\noindent
{\bf Acknowledgements:} This work has been carried out thanks to the support of the OCEVU Labex
(ANR-11-LABX-0060) and the A*MIDEX project (ANR-11-IDEX-0001-02) funded
by the "Investissements d'Avenir" French government program managed by
the ANR.

\end{document}